\documentclass{aa}  

\usepackage{graphicx}
\usepackage{txfonts}
\usepackage{natbib}
\bibpunct{(}{)}{;}{f}{}{,} % to follow the A&A style

\newcommand{\ms}{\,$\mathrm{m\,s^{-1}}$}

\begin{document}

  \title{The Be-test in  the Li-rich star \#1657 of  NGC~6397:\\ evidence for Li-flash in RGB stars?
  \thanks{Based on observations collected at ESO, VLT, Chile, Proposal 091.D-0198(A).} }

  %\subtitle{LETTER TO THE AUTOR}

   %\author{L. Pasquini \inst{1}
   %       \and
   %       L. Pasquini\inst{2}\fnmsep\thanks{Just to show the usage
   %       of the elements in the author field}
   %       }

   %\institute{Institute for Astronomy (IfA), University of Vienna,
    %          T\"urkenschanzstrasse 17, A-1180 Vienna\\
    %          \email{abrucala@mpe.mpg.de}
    %     \and
    %         ESO, Garching, ...\\
    %         \email{lpasquin@eso.org}
        %     \thanks{ciao}
    %         }
    
  \author{L. Pasquini\inst{1} \and   A.  Koch\inst{2} \and R. Smiljanic\inst{3} \and P. Bonifacio\inst{4} \and A. Modigliani\inst{1}   }

   \institute{ESO -- European Southern Observatory, Karl-Schwarzschild-Strasse 2, 85748 Garching bei M\"unchen, Germany
   \and Zentrum f\"ur Astronomie der Universit\"at Heidelberg, Landessternwarte, K\"onigstuhl 12, 69117 Heidelberg, Germany 
   \and Department for Astrophysics, Nicolaus Copernicus Astronomical Center, ul. Rabia\'nska 8, 87-100 Toru\'n, Poland
   \and  GEPI, Observatoire de Paris, CNRS, Univ. Paris Diderot, Place Jules Janssen 92190 Meudon, France 
    }          
             
  % \date{Received September 15, 1996; accepted March 16, 1997}
    \date{Received  / Accepted }

% \abstract{}{}{}{}{} 
% 5 {} token are mandatory
 
  \abstract
   {The Li-rich turn-off star recently discovered in the old, metal-poor globular cluster NGC~6397 could represent the smoking gun for some fundamental, but very rare episode of Li enrichment in globular clusters and in the  early Galaxy. }
   {We aim to understand the nature of the Li enrichment  by performing a  spectroscopic analysis of the star, in particular of its beryllium (Be) abundance, and by investigating its binary nature.}
   {We used the VLT/UVES spectrograph  to observe the near UV  region where the \ion{Be}{ii} resonance
doublet and the NH bands are located. We also re-analyzed the Magellan/MIKE spectra
of Koch et al. for C and O abundance determination.
}
   {We could not detect the \ion{Be}{ii} lines and derive an upper limit
of log (Be/H)$< -12.2$, that is consistent with the Be observed in other stars of 
the cluster. We could detect a weak G-band, which implies
a mild carbon enhancement [C/Fe]$+0.4\pm0.2$. 
We could not detect the UV NH band, showing that this star
is less N-enhanced than other stars of the cluster, and we derive
an upper limit [N/Fe]$< 0.0$. For oxygen we could not  convincingly detect
any of the near UV OH lines, which implies that oxygen
cannot be strongly enhanced in this star. This is consistent with the
detection of the strongest line of the \ion{O}{i} triplet at 777\,nm,
which is contaminated by telluric absorptions but is consistent with
[O/Fe]$\sim 0.5$. Combining the UVES and Mike data, we could not detect
any variation in the radial velocity greater than 0.95 kms$^{-1}$ over 8 years.}
{The chemical composition of the star strongly resembles that of `first generation' NGC6397 stars, with the huge Li as the only 
deviating abundance. 
Not detecting Be rules out two possible explanations of 
the Li overabundance: capture of a substellar body and
spallation caused by a nearby type II SNe. 
%Two of the other proposed
% scenarios, accretion from an AGB or RGB companion, have to account
% for the lack of detection of radial velocity variations. This is possible
% if the orbit is seen almost face-on, the period is very long or the binary
% has since been disrupted. Accretion during a close encounter with an unbound star
% is another possibility. The AGB scenario may have the further difficulty that it
%must account also for the lack of enhancement in Na.
Discrepancies are also found with respect to other accretion scenarios,   except for contamination by the ejecta of a star that has undergone 
the  RGB Li-flash. This is at present the most likely possibility for explaining the extraordinary Li enrichment of this star.  }


   \maketitle
%
%________________________________________________________________
%
%
\section{Introduction}
%
%
%   The study of peculiar objects can be of enormous interest;  
% extremely metal poor stars are a great example:  in spite of the fact that 
% the Halo is a tiny fraction of the mass of the Galaxy, and the low metallicity tail is a tiny fraction of the halo, 
% the discovery of  few  extremely  metal poor objects 
% can provide direct information on the properties of the First Stars (see, e.g. Bonifacio et al. 2012 and references therein). 
%
Rare, peculiar objects in the Milky Way halo and in globular cluster (GCs)  can be extremely interesting, because  they may be  smoking 
guns for understanding  phenomena that occurred in the early Galaxy. The extremely Li-rich star
 \#1657 in NGC~6397 ($\equiv$2MASS J17410651-5342290) is one of these objects. Lithium is a complex element because it can be created in a number of  
 ways such as Big Bang nucleosynthesis, in the cool bottom burning process during the asymptotic giant branch (AGB) phase, or via spallation.
On the other hand, it is easily destroyed in the stellar interiors. 
 In the context of GCs, Li is fundamental for constraining cosmological models, understanding the role of 
 diffusion, and  explaining the nature of the multiple populations now clearly detected in most clusters \citep[see][for a review]{Korn12}. 

NGC~6397 is one of the best-studied GCs. It is metal-poor ([Fe/H] $\sim -2$ dex) and
has been age-dated through its Beryllium abundance and stellar evolutionary models (Pasquini et al. 2004). 
In this metallicity range, warm stars around the turn off (TO)
display a constant lithium abundance, irrespective of their temperatures, around log (Li/H)+12 = 2.2, the 
so-called ``Spite plateau'' (Spite \& Spite 1982). 
Extensive spectroscopic surveys have shown that  turn-off (TO) stars in NGC~6397 
have a Li abundance very close to the ``Spite  plateau''  (Th\'{e}venin et al. 2001, Bonifacio et al. 2002; Lind et al. 2009; 
Gonz\'alez Hern\'andez et al. 2009), and a debate is  going on about whether the Li abundance  
around the TO shows the signature of diffusion (Korn et al. 2006). 

Star \#1657 is located at the cluster  TO. While it is fully representative of the remainder of the GC giant and TO stars in terms of metallicity,  
$\alpha$-element abundance ratios, and stellar parameters, it has  a Li abundance about 100 times higher than any   
other NGC~6397 stars. With an exceptional abundance  of 
log($^7$Li/H)+12 = 4.2 (assuming non-LTE),  it is  one of the stars  with the highest Li ever observed (Koch et al. 2011, 2012), and
only very few other dwarf or giant stars show comparable levels of enhancement (Reyniers \& Van Winckel 2001; Deliyannis et al. 2002; Monaco et al. 2012; Adam\'ow et al. 2012).
{\em What produced such a high Li abundance in a TO star of a GC?} 
Considering that the primordial abundance of Li produced in the Big Bang is almost 100 times lower than the one observed in \#1657,   some 
extra production and contamination mechanism must be invoked, either by engulfing substellar objects (see, e.g., Pasquini et al. 2007), 
or by the mass transfer from a more massive star that went through an evolved phase, nurturing the manufacturing of extra-Li. The best contenders for the latter   
are AGB stars (Cameron \& Fowler 1971) or the acting of the cool bottom process (CBP) on the red giant branch (RGB; Sackman \& Boothroyd 1999).
Alternatively, Li could have been 
produced by the explosion of a nearby supernova (SN) through spallation 
% $\nu$-process in the SNe?)}
(cf., e.g., Smiljanic et al. 2008). 

All of these processes  have some serious shortcomings, as discussed in Koch et al. (2011, 2012). 
However, one key test  can be performed to easily distinguish the main scenarios, viz.  the beryllium test.
Amongst the different hypotheses,  the mass transfer from an evolved
companion (of any kind) cannot add any Be to the star, since Be is  easily destroyed in the stellar interiors. 
On the other hand,  either the engulfing of a substellar companion 
or the spallation produced by a nearby  SN explosion would enhance Be in the atmosphere of the presently observed star by a large factor.

The engulfing of a substellar companion with the same (Li/Be) ratio as in the primordial GC material would enhance the Be abundance by a 
factor of $\sim$65 with respect to the original Be of the star (i.e., the same overenhancement as seen for Li). 
When there is an energetic, near-by event, such as 
the explosion of a super- or hypernova, the  enhancement factor can reach as high as $\sim$5600 
(assuming a ratio of Li/Be = 4.2, by number,  in the spallation material as suggested by the  
hypernova models of Nakamura \& Shigeyama 2004). 

% Figure 1 shows the simulation of the 
% spectrum of star \# 1657 
% with log(Be/H) $\sim$ $-$10.50 
% (an enhancement of $\sim$ 65 times in comparison with 
% the expected Be abundance of 
% a normal star in NGC~6397). It is clearly possible 
% to detect the line even with a S/N ratio of 
% 15 at the UVES resolution. A comparison with the other 
% TO stars (Pasquini et al. 2004)
% will immediately show the Be enhancement. 
% Star \# 1657 provides therefore  an extraordinary opportunity to learn about 
% Li production in GC. 

We consider that, as  for other GCs, NGC~6397 shows 
evidence of multiple populations, both in its color-magnitude diagram (di Criscienzo et al. 2010) and in chemical abundance space, 
with stars having the same metallicity but different light element abundances 
(e.g., Carretta et al. 2009; Lind et al. 2011; Gratton et al. 2012, and references therein). 
In this context, Shen  et al. (2010) demonstrate for the more metal-rich GC 
NGC~6752 that the first-generation stars must have enriched 
some of the  second-generation stars in Li. Similarly,  in NGC~6397, Pasquini et al. (2008)
find a  Li-rich but O-poor star,  which also requires some  pollution with Li by a prior population. 
%

%The spectra of Koch et al. (2011a,b) only permitted the measurements of Fe, Na, and $\alpha$-element abundances so that 
%neither oxygen, carbon, nor  nitrogen abundances are known for star \#1657. 
%Koch et al. (2012) note a low [Na/Fe] ratio, but without any available measure for O, this object could not be unambiguously assigned to any 
%stellar generation within NGC~6397 nor could it be tested whether it has been contaminated by CNO-processed material, 
%{\bf which would be indicative of...}.
%
%Thus we obtained new spectra of this peculiar object that allowed us to expand our chemical knowledge by measuring more elements.  
%
The present work aims to investigate the Be abundance and that
of other elements, to better characterize the properties of this star.
\section{Observations \& data reduction}
The observations were obtained with the UVES spectrograph at the VLT (Dekker et al. 2000). The bluest setup,  centered at  346\,nm was used to cover the UV part of the spectrum, together with the red setup 
centered at 580 nm. Seven blue observations of 1.5 hours each were requested  and executed in the period of April 14 -- June 18, 2013. For each UV observation, two red observations were acquired.
%, and all  are available in the ESO archive.  
Two observations were repeated because they were hampered by poor conditions, so the total number of observations consists of 17 red  and 9 blue spectra. 
The blue slit was set at a width of 1.1$\arcsec$, yielding a resolving power of R$\sim$38000, while  the red slit width was selected at 0.6$\arcsec$,  corresponding to R$\sim$67000.

The spectra were reduced using the ESO reflex pipeline (Larsen et al. 2008; Ballester et al. 2011).  To gather a sufficiently high signal-to-noise (S/N) ratio in the extreme UV, where the Be lines are located (around 313 nm), the blue spectra were obtained  with the detector in binned mode. The extracted, wavelength-calibrated spectra were  corrected for earth motion and finally combined. 
Out of the available nine blue spectra, two revealed a S/N ratio that was too low to be used, and one was contaminated by a very strong, high-energy event (cosmic ray) 
right in the region of the Be lines, and they have not been used. 
The remaining six spectra were combined to form the resulting  spectrum used in the Be analysis of star \#1657. 
%
%
% {\bf Notes on reduction of red spectra?}
%
%
\section{Analysis}
\subsection{Radial Velocity}
As already noted by Koch et al. (2011),  the spectrum of star \#1657 is contaminated by the flux of a second star with a very  different radial velocity. 
We have used two ways of estimating radial velocities: 
For the GC TO primary, we directly used the Doppler-shift of the strong and deep Li resonance line at 670.7 nm. 
For the contaminating foreground companion, we cross-correlated the 
lower part of the red spectrum  with a template of a solar-type star (Melo et al.  2001). 

In the resulting cross-correlation function (CCF), the peak of the contaminant  is very pronounced, 
indicating that this object is much more metal rich than a typical member star of NGC~6397. 
The peak of star \#1657 itself is also often present in the CCF, but with a weak and more uncertain CCF peak, 
when compared to the direct measurement of the Li line. 
Moreover, the Li line is visibly strong and measurable in all spectra, so that we adopt  these measurements for the remainder of the paper. 

The radial velocity variability of the GC star  from the Li line on the 17 exposures is  350\ms~(1$\sigma$) and thus perfectly compatible with 
instrumental and centering instabilities. 
Exposures taken over the same night show a 1$\sigma$-scatter of only 220\ms~, and  these have been averaged in Table~1. 
%
%
%\small
\begin{table}[htb]
\caption{Radial velocities for star \#1657 (first two columns), measured by two methods, and the contaminant. See text for details.}
\centering
\small
\begin{tabular}{lccc}
\hline\hline
  &  \multicolumn{3}{c}{v$_{\rm HC}$ [km\,s$^{-1}$]} \\ 
\cline{2-4}
\raisebox{1.5ex}[-1.5ex]{Date of obs.} & Li\,670.7 nm &   CCF &   Contaminant  \\
\hline
2013 04 27   & 19.37  & 20.2  & $-$101.2 \\
2013 04 29   &  19.06  & 18.8  & $-$102.8 \\
2013 05 27   &  19.26  &   \dots        &   $-$101.2 \\
2013 05 28   &  19.07  &   \dots         &   $-$102.6  \\
2013 06 15   &  19.39  &  19.9 &  $-$101.8 \\
2013 06 16   & 18.80     &     \dots       &  $-$102.2  \\
\hline
\end{tabular}
\end{table}

% We note that the CCF for the primary \#1657  shows a larger FWHM compared to the metal-rich contaminant (11.4 \kms~ vs. 10.2 \kms), which could indicate % possible rotation and we will come back to this point 
% in Section~{\bf xy}. 
%
%
%
%
Figure~1 shows the radial velocity measurements of the two components when combining 
the previous observations obtained with the Magellan/MIKE instrument (Koch \& McWilliam 2011; Koch et al. 2011, 2012) with our new ones by UVES. 
To aid the readability of the velocity curves, we split the time axis so as to shorten the gaps of up to six years between the observing runs. 
\begin{figure}[htb]
\centering
\includegraphics[width=1\hsize]{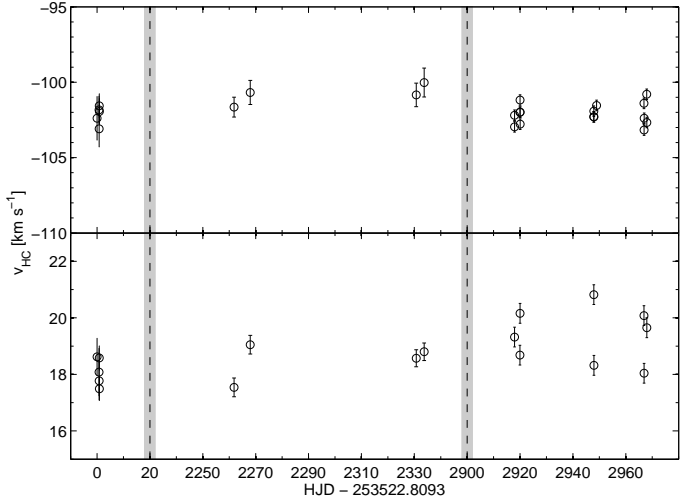}
\caption{Combined radial velocity of star \#1657 (bottom panel) and the metal-rich foreground contaminant (top panel). Segmented range of the x-axis covers more than 8 years of observations. 
The vertical bars separate the three observing runs. }
\end{figure}

We note that there are systematic differences in the observational setups and in the data analysis of the MIKE and UVES observations that could lead to potential biases. 
%While the latter used a slit width of 0.5$\arcsec$ and obtained a resolution of R$\sim$40000, the present (red) setup achieved  a higher resolving power. 
%Furthermore, Koch et al. (2011a) obtained their CCF by using a red-giant template spectrum, which clearly showed the secondary foreground peak, while 
% putting more emphasis on the GC TO primary star. 
It is then striking that there is no evidence of any significant, systematic velocity variation within the combined data. 
In fact, we could not find any suitable orbital solution that would describe the observed velocity data of either component. 
We conclude that the star does not show any radial velocity variations larger than  0.95 km\,s$^{-1}$ (1$\sigma$)
over eight years of observations, and its velocity is in perfect agreement with the estimated mean systemic velocity of the cluster (Milone et al. 2006).
%

%While we cannot exclude that the motion of the contaminant star is variable, the radial velocity of the star \#1657 is perfectly stable over the entire observed period of 8 years, 
%to the best of 0.95 km\,s$^{-1}$ (1$\sigma$) {\bf  or which measure should we use to make the point of non-variablilty?}. 
%Moreover, as mentioned above, there is no evidence for short-period variations within single nights, nor within the individual (weeks to months) runs.  
%Thus the TO star does not show any significant evidence for binarity.  

%We note in passing that also any radial velocity  variability of the contaminant star may not prove that it is in a binary. The 
%star-slit geometry will vary for our observations (both for the UVES and the MIKE data), because the slit orientation follows the parallactic angle. 
%Given that the superposition of the contaminant with star \#1657 is a geometrical one this effect will likely induce some variability 
%in the observed velocities of the star only due to the geometrical  shifts of the contaminant star in the slit.  
%We finally note that he mean radial velocity of star \#1657 is in perfect agreement with the estimated mean systemic velocity of the cluster (Milone et al. 2006).
%
%
%
\subsection{Be abundances}
We adopt the stellar parameters derived by Koch et al. (2011), namely T$_{\rm eff}=6282\pm250$ K, 
log\,$g=4.1\pm0.2$ dex, $\xi=1.2\pm0.8$ km\,s$^{-1}$, and an iron abundance, measured from the neutral species, of [Fe/H]=$-1.93\pm0.06$(stat.)$\pm0.18$(sys.).
The Be doublet region is shown in Figure 2, with  the synthetic spectrum overimposed, computed as in Smiljanic et al. (2011). 
The stellar spectrum has a low  S/N $\sim 10$ and shows no clear detection of the Be lines. 
We determine an  upper limit of log(Be/H) = --12.2  by performing Monte Carlo simulation on synthetic
spectra with different abundances and trying to detect the \ion{Be}{ii} lines by basic statistics (average counts, smallest flux pixel,
highest flux pixel). 
This upper limit is compatible with the Be observed in the other two TO stars of NGC~6397 for which this line has been detected (Pasquini et al. 2004).
A strong conclusion is that star \#1657 does not show any evidence of Be enhancement with respect to stars of the same cluster or with respect to field 
stars with similar metallicity. 
\begin{figure}[htb]
\centering
\includegraphics[width=1\hsize]{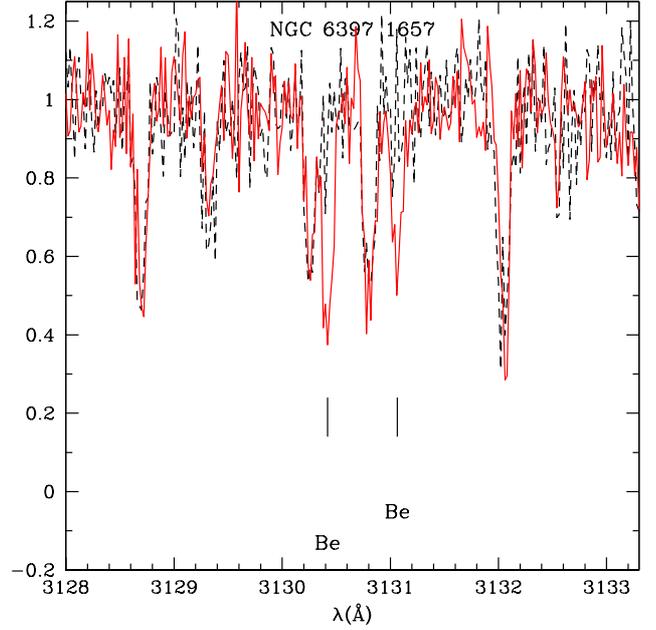}
\caption{Be-lines in \#1657. The observed spectrum is shown as black, dashed line. The red line is a synthetic spectrum with Be/H=-11.5, to which noise has been added. Both Be lines are not detected and only a conservative upper limit could be set  at  log(Be/H)=-12.2 .}
\end{figure}
%

%{\bf Following from Rodolfo's last email}. 
When assuming 
the linear relation derived in Smiljanic et al. (2009) between [Fe/H] and log(Be/H), star \#1657 should have log(Be/H) $\sim  -12.8$.
Assuming that the Li enhancement is due to pollution,  that initial Li should be A(Li) = 2.20 (Spite \& Spite 1982; Lind et al. 2009, Gonz\'{a}lez Hern\'{a}ndez et al. 2009), and  that the polluting material had normal
Be/Li composition (equivalent to A(Li) = 2.20 and log(Be/H)$\sim -12.8$), the star should have accreted an amount of Be that would increase its chemical composition to log(Be/H)$\sim ~ -11.5$. Assuming the Be abundance from the field star regression (log(Be/H)=-12.8) as reference value is a conservative assumption, because this value  is lower than what is observed in the TO  stars of the cluster. 
% -- I'd need to double check it, because I think I did some circular reasoning in my calculations.
%Anyway, then I calculated 2 spectra with these Be abundances and added noise to simulate S/N = 10.
%So if the numbers above are not wrong by too much, our upper limit will be good enough (more than good actually) to exclude that the polluting material had Be in normal quantities.
%
%
%\subsection{Light element abundances}
%
%
%\begin{center}{\em CNO}\end{center} 
\subsection{C, N, and O}
The UVES spectra do not cover any region that can
be used to determine the carbon abundance. The Mike spectra
cover the G-band, albeit at low S/N. In such warm dwarfs the
G-band is very weak, but it is clearly detected. We conclude
that carbon is mildly enhanced in this star: [C/Fe]$=0.4\pm 0.2$. 

An excellent diagnostic for nitrogen is provided by the UV NH band
and 336\,nm, whicht has been used by Pasquini et al. (2008) to measure
N in three TO stars in this cluster. 
To our surprise we were unable to detect the band in 
star \#1657, whose spectrum in the NH band region is distinctly
different from that of star A228, which has very similar
atmospheric parameters and for which Pasquini et al. (2008)
measured [N/H]=--0.75. In fact,  all  three stars in this
clusters measured by Pasquini et al. (2008) show similar N enhancement
with respect to iron. 
We estimate [N/Fe]$< $--2.0, both from comparison with synthetic
spectra and differentially with respect to star A228.

To search diagnostics for the oxygen abundance we
inspected the UV spectra for the presence of OH lines
(see Gonz\`{a}lez Hern\`{a}ndez et al. 2010), but we were
not able to detect any. The same is true 
for the three stars observed by Pasquini et al. (2008),
this allows us to rule out a strong enhancement 
in oxygen in this star, with respect to other stars
in the cluster.  
This result is also consistent with the detection
of only  the strongest line of the \ion{O}{i}  
triplet at 777.1\,nm in the Mike spectrum.
We have a very uncertain estimate of the  equivalent width 
of about 1.9 pm, which corresponds to [O/Fe]$\sim 0.5$ when 
taking  the NLTE corrections
for oxygen into account, according to the prescriptions of Shen et al. (2010).  
The line is possibly contaminated by telluric absorption
on its red wing. Comparison of the line in this star with
other TO stars of the same cluster observed under the same
conditions with MIKE shows the strength of the 
\ion{O}{i} triplet in star \# 1657 to be quite typical.

%
% 
%\begin{center}{\em Sodium}\end{center} 
\subsection{Sodium}
Koch \& McWilliam (2011) obtained $\alpha$-element abundance ratios of three red giants and two TO stars in NGC~6397 and  could measure Na-abundances in the three red giant stars of their sample. 
Upon revisiting the MIKE spectra, here we also measured equivalent widths (EWs) of the near-infrared Na line at 819.5 nm  of the TO stars. 
As a result, we obtain [Na/Fe] ratios of  $-0.03$ and $-0.40$ dex for the two regular (i.e., Li-normal) TO stars and, for the Li-rich \#1657,  an upper limit of [Na/Fe]$< -0.56$, based on an EW of $<$1.6 pm. 
We can use this to associate the giants and one of the TO stars with the second generation of stars in the GC, characterized by elevated Na- and depleted O abundances (Carretta et al. 2009). 
The very low [Na/Fe] ratios  in the remaining TO star and 
the present \#1657 suggests that they are part of the first stellar generation of NGC~6397, which formed 
from material that was not yet enriched with 
Na-rich ejecta coming from p-capture reactions in early, massive AGB-polluters (e.g., Decressin et al. 2007; D'Ercole et al. 2008). 

%Lind et al. (2009) trace a moderate  Na-Li anti-correlation in a large sample of stars in NGC~6397, which is connected to the occurrence of at least two generations of stars in this GC. 
%Since also our star in question is very low in Na and, in turn, very rich in Li, this enhancement could be similarly related to the multiple generations seen in NGC~6397;
% however the detailed evolutionary mechanisms have yet to be identified. 

%
%
%
%
%
%
\section{Discussion}
Table 2 summarizes the abundances measured in star \#1657, compared with the value range observed in first-generation NGC6397 stars.
 Exceptions are obviously Li overabundance and possibly  N  underabundance. 
 The abundance of N in TO stars of NGC6397 seems highly variableI and apparently independent of any other light element abundance. Pasquini et al. (2008) find high N in three TO stars irrespective of their light element  abundances, and Carretta et al. (2005) show that N can span almost two orders of magnitude among NGC6397 subgiants, with two N-poor stars  among the  richest in  O, but all span a very narrow C abundance range.  If we assume an analogy with field stars, we expect a low N abundance for the first generation of cluster stars.  
 Chemically therefore  star \#1657 has no signature of pollution from processed material; in contrast all its values point toward an NGC6397 first-generation star, with the exception of Li.

Before the present study, Koch et al. (2011, 2012) discussed five 
possible causes of the Li- pollution for star \#1657: 
({\sl i})
capture of a substellar body; 
({\sl ii})
type II Supernovae;
({\sl iii})
diffusion;
({\sl iv})
contamination by an AGB companion;
({\sl v})
contamination from an RGB 
companion having undergone cool bottom processing.

\begin{table}[htb]
\caption{Elemental abundances for star \#1657 and for typical first-generation NGC6397 star.}
\centering
\small
\begin{tabular}{lll}
\hline
%   &  \multicolumn{3}{c}{v$_{\rm HC}$ [km\,s$^{-1}$]} \\ 
% \cline{2-4}
%\raisebox{1.5ex}[-1.5ex]{Date of obs.} & Li\,670.7 nm &   CCF &   Contaminant  \\
Element & \#1657 & NGC 6397 \\
\hline
[Fe/H]    &          $-$1.93 $\pm$0.2       &      $-$2.02  \\
Li/H       &           4.2                           &        2.1/2.2\tablefootmark{d} \\
Be/H     &     $<$ -12.2                    &     $-$12.2/$-$12.3\tablefootmark{a} \\
{[C/Fe]}   &     0.4$\pm$0.2       &        $<$0.5\tablefootmark{b}   \\
{[N/H] }   &      $<-2$                      &    $-$3 / $-$1\tablefootmark{b,c}   \\
{[O/Fe]}   &         $\sim$+0.5               &     +0.4  / +0.6\tablefootmark{b}  \\
{[Na/Fe]}  &          $<-0.56$          &   $-$0.35  / $-$0.65\tablefootmark{d}    \\
\hline
\end{tabular}
\tablefoot{
\tablefoottext{a}{Pasquini et al. (2004)}
\tablefoottext{b}{Carretta et al. (2005)}
\tablefoottext{c}{Pasquini et al. (2008)}
\tablefoottext{d}{Lind et al. (2009)}}
\end{table}

As far as spallation, planet engulfment, and internal mixing/diffusion are concerned,  
the first two can be excluded by the regular abundance of Be. Internal mixing and diffusion can be  excluded because of  the  extremely high abundance of Li observed. The  two last processes implying pollution from an evolved companion  remain, in principle,
viable. 

If Li has been produced in a companion then  our radial velocity measurements coupled with those of 
Koch et al. (2011) imply that the orbit is highly inclined or has a very long period
(hundreds or thousands of years) and that the companion
star should now be subluminous (e.g., a white dwarf).
Alternative explanations could be that the binary
has since been disrupted, e.g. due to collisions
or tidal effects, or that the accretion took place
during a close encounter of two unbound stars. Of course it cannot be excluded that the stars formed from a polluted gas, therefore not only the external atmosphere  
composition has been altered. This would solve the problem of the missing companion, but would imply an enormous amount of Li, because in this case the whole mass of the star would have been affected.  

Pollution through the winds of a companion that passed through a (super-) AGB phase, during which the extra-Li would have been 
produced through hot bottom burning (Cameron \& Fowler 1971; Ventura \& D'Antona 2011) is in principle a possibility. In the lower mass regime, this could lead to elevated levels of $s$-process elements, which is not observed in the Li-rich star \#1657 (Koch et al. 2011). %; Sect.~3.4. below {\bf (tbc)}). 
However, the absence of significant $s$-process-rich material cannot be unambiguously used as an argument against AGB contamination, since 
that material will not be dredged-up from deeper layers in the more massive (6--8 M$_{\odot}$) {\em super-}AGB stars (Ventura \& D'Antona 2010). 
Nevertheless, a super-AGB star  should also produce sizeable amounts of the light elements Na and K, none of which are seen in \#1657. 
This means that we can reject the hypothesis of the extra lithium in this star originating in a super-AGB envelope.
 
%  The  fact that the star has a composition of a 'first generation' cluster star  together with the huge Li,  points against a simple pollution scenario where polluted and pristine material are mixed. 
 
Pollution by an RGB star does not contradict  the observations at our disposal, if one would admit the presence of a  `Li flash'  in RGB stars.  
Lithium-rich giants are known in the Galaxy (de la Reza 1996), and they present in external dwarf galaxies at the rate of 1$\%$ (Kirby et al. 2012). Like star \#1657, they seem to  only be enriched in Li. 
They possibly represent a short phase of stellar evolution, followed by a Li flash (de la Reza et al. 1996). No RGB star with such  high Li content has been discovered so far, and even if several propositions exist in the literature (Denissenkov \& Weiss 2000, Palacios et al. 2001),  
no detailed mechanism has been proven for such a flash.  
 But in first approximation, this would fit with the absence of enhanced Na, the low N, and the composition of a first-generation cluster stars  not showing evidence of binarity.  In this sense it would be  different from what produces the well-known light elements anti-correlations in globular clusters.  
The high abundance of C observed may contradict the trend of diminishing  C abundance along the RGB (Carretta et al. 2005), but given that we do not know when and how the hypothetical Li flash occurs, this discrepancy may not be so stringent. Also, the sudden extra mixing invoked to explain the 
Li flash is of a substantially different timescale than the slow extra mixing invoked along the RGB to explain the slow decay of C with luminosity.
We finally notice that, if Be/H was confirmed at a level comparable to the other NGC6397 stars (Pasquini et al. 2004), then the RGB pollution hypothesis should also be dropped, since Be has  already been diluted in the atmospheres of evolved stars. 

Exploring  the RGB hypothesis further, de la Reza et al. (1997) argue that, at a certain point in the life of a giant, an abrupt mixing mechanism can lead to a rapid surface injection of material with fresh internal $^7$Be, which is  rapidly
transformed to $^7$Li. The same mechanism would then enable  the formation of a circumstellar shell of gas and dust that is later ejected in the interstellar medium.  
Denissenkov \& Herwig (2004) show that extra mixing, driven by faster rotation, is able to produce a Li flash. 
They argue that such a spinning could be produced in close binaries or by engulfing a massive planet. Modeling the Li flash is beyond the scope of this paper. We note, however, that an alternative source of extra mixing could come from the possible coupling between shells and cores in giants, given the recent finding by the Kepler satellite that cores of RGB stars rotate $\sim$ 13 times faster than the envelopes (Beck et al. 2012, Deheuvels et al. 2012). The core rotation is faster than what
 is expected by solid rotation, but slower than expected if shellular rotation is assumed, indicating that  core-envelope coupling is present at a certain level. 

Although  the presence of a Li-O correlation, as present in NGC\,6752,
has not been established in NGC 6397, we should expect a very
large oxygen abundance if such a relation  existed and if 
 the extremely high Li abundance
in star \#1657 was linked to the phenomenon
producing that correlation.
However, a strong enhancement in O is firmly ruled out by our observations.

Exotic explanations, such as  Li production around
compact objects, which could explain the high
Li abundances observed in low-mass X-ray binaries
\citep[see, e.g.,][]{guessoum} seem unlikely,
since Casares et al. (2007) favor the  simpler Li-preservation
scenario due to tidal locking rather than the Li production in these systems 
for the case of  Cen-X4 .

The Li abundance in \#1657 is most likely not directly
linked to the spread in Li abundances observed
in NGC\,6397 and other GCs, but is linked to 
a very exceptional event. Contamination by an RGB that went through the Li-flash  so far seems the most likely possibility.  Regardless of the explanation,
it shows that local production of Li is possible in GCs, with peaks well exceeding the cosmic values.
This work reinforces the concept that extreme care should be used when inferring the properties of primordial nucleosynthesis based on observations of Li in GCs -- 
not only because of the clear presence of second-generation stars that show light elements correlations and that can be Li polluted, 
but also because even stars that can be considered bona fide first-generation may be polluted in Li, the RGB Li-flash was a common phenomenon. 
If every RGB star underwent the Li-flash, as argued by de la Reza et al.,  we could predict that the absolute Li abundance and its spread among 
GC TO stars (even selecting only first-generation stars) 
should be higher than what is observed among field stars, because of the presence of  `disguised' second-generation stars that,  like star \#1657, have been  only contaminated  by the ejecta of RGB stars.

\begin{acknowledgements}
LP acknowledges the Visiting Researcher (PVE) program of the CNPq Brazilian Agency, at the
Federal University of Rio Grande do Norte, Brazil.      
AK acknowledges the Deutsche Forschungsgemeinschaft for funding from  Emmy-Noether grant  Ko 4161/1. 
PB acknowledges support from the Conseil Scientifique de 
      l'Observatoire de Paris and from the Programme National
      de Cosmologie et Galaxies of the Institut National des Sciences
      de l'Univers of CNRS.RS is supported by the National Science Centre of Poland through grant 2012/07/B/ST9/04428.
\end{acknowledgements}

%-------------------------------------------------------------------

%-------------------------------------------------------------------------------

% \bibliographystyle{aa} % style aa.bst
% \bibliography{ pippo} % your references Yourfile.bib

\end{document}